**Title:** Antibiotic resistant characteristics from 16S rRNA

**Author:** Casey R. Richardson

**Affiliation:** Morton Bioinformatics, Morton, Texas, United States of America

**Abstract**

**Background:** Microbiota have evolved to acclimate themselves to many environments. Humanity is become ever increasingly medicated and many of those medications are antibiotics. Sadly, Microbiota are adapting to medication and with each passing generation they become more difficult to subdue. The 16S small subunit of bacterial ribosomal rRNA provides a wealth of information for classifying the species level taxonomy of bacteria.

**Methodology/Principal Findings:** Experiments were collected utilizing broad and narrow spectrum antibiotics, which act primarily on DNA. In each experiment a statistically significant, unique and predictable pattern of sequential and thermodynamic stability or instability was found to correlate to antibiotic resistance.

**Conclusions/Significance:** Classification of antibiotic resistance is possible for some species and antibiotic combinations using the 16S rRNA sequential and thermodynamic properties.

## Introduction

Extremophiles thrive in uncommonly harsh environments. For example, thermophiles live in temperatures ranging from 45 to 122C [1] [2] [3], halophiles dwell in environments with high salt content [4], and most importantly antibiotic resistant life forms which flourish despite antibiotics. Antibiotics resistance is an important area of study because many diseases are developing resistance to antibiotics. I hypothesize that Prokaryotic extremophiles can characterize by their 16S rRNA sequential characteristics. Features I have examined to justify this hypothesis are nucleotide and dinucleotide frequencies, adenosine and uracil (A+U) and guanine and cytosine (G+C) content, adjusted base pairig propensity $P_b$, adjusted base pair distance $d_D$, adjusted Shannon entropy $d_Q$, and minimum free energy (mfe) [5]. Secondary structure and G+C content have been shown to characterize the living environment of bacteria [6]. 16S rRNA have also been shown to play a part in a bacteria's fitness [7].

One recent study has been performed on the human gut's microbiota adaptation to Ciprofloxacin (cp), a synthetic antibiotic used to treat bacterial infections [8]. Another study has been performed on the swine gut's microbiota in the presence of ASP250 (chlortetracycline, sulfamethazine, and penicillin) and Carbadox [9]. A third study was performed on the microbiota of mice in the presence of Ampicillin, Vancomycin and an antibiotic cocktail composed of Metronidazole, Neomycin, and Vancomycin (MNV) [10]. Neomycin is an aminoglycosides but the other antibiotics in the studies are not. Aminoglycosides are known to have a 16S resistant pathway conferred by methyltransferases [11-14]. Study of antibiotic resistant bacteria has focused on lateral gene transfer, which confers antibiotic resistance but most have not focused on genera or species-level determinants observable from 16S rRNA [15-19]. I conclude that base pairing and dinucleotide frequencies in the 16S rRNA characterize species of bacteria that are able to withstand specific antibiotics. This paper explains a method of predicting broad-spectrum antibiotic resistance to help facilitate a transition to targeted (even if remaining broad-spectrum) antibiotic therapy [20].

## Materials and Methods

rRNA data was provided by the SILVA database. Pyroseqencing data was obtained from NCBI SRA and by direct contact with authors. Each read was clustered into an operational taxonomic unit (OTU) at a percent id of 97% using USEARCH. Because pyrosequencing data results in reads of relative abundance, each sample's run was rarefied by dividing by the total number of reads for each individual sample. Percentage of representation allows comparison between samples of unequal total abundance. Each OTU was mapped to the nearest SILVA full length 16S sequence [21] [22]. For comparison purposes, each OTU of the PNAS data was also mapped to the full Greengenes database and the results were equivocal. VienneaRNA was used to collect thermodynamic properties of 16S rRNA. genrnastats.pl and genRandomRNA.pl written by Stanley NG Kwang Loong were modified to provide all the characteristics and random negative rRNA to test the multiple linear regression. [23] Multiple linear regression was performed with the R programming language. Pearson's correlation coefficients, binomial distributions and t distributions were calculated using Microsoft Excel.

## Results and Discussion

I examined the thermodynamic stability of 127,524 genes from Greengenes database and found that 16S rRNA could be used to classify extremophiles based on their maximum free energy and normalized maximum free energy. [22] Thermopiles have highly stable rRNA thermodynamic properties. [24,25] I add from my observation that halophiles also have a highly stable secondary structure and phytoplasma have a very weak secondary structure. Phytoplasma have no cell wall and must adapt to the various situations they are presented with; this ability is reflected in their unstable rRNA.

Concluding that secondary structure could be used the classify extremophiles, I analyzed the pyroseqencing data from the human gut's microbiota taken in the presence or absence of cp. Ciprofloxacin is a broad-spectrum quinolones that inhibits separation of bacterial DNA. In "Incomplete Recovery and Individualized Responses of the Human Distal Gut Microbiota to Repeated Antibiotic Perturbation" (PNAS), 30 samples were acquired before, during, or after cp administration and 8 samples were taken during the regiment of cp. [8] My analysis focuses on the total samples taken and the samples taken during cp. Both datasets were analyzed as a whole using Pearson's correlation coefficient with a cut off value of $p < .05$.

Table 1 shows PNAS intrinsic 16S RNA thermodynamic characteristics. Three classes were chosen to represent the biosphere of the human gut with and without the selective pressure of an antibiotic. Total expression is the sum of all the rarefied sequences from 16S reads for all individuals involved in the experiment. I considered the collective group of patients as one biosphere under equal selective pressure from a single antibiotic. Total resistance is the sum of all the rarefied of an OTU's 16S reads for all samples taken during antibiotics for all individuals. Average percent resistance is the 16S rarefied reads taken during antibiotics divided by total amount of rarefied reads averaged for the three individuals. Finally, a simple percent resistance was calculated dividing the rarefied total expression and total resistance. The percent resistance classes show which bacteria thrived under the selective pressure of antibiotics. A biosphere approach to the human gut is made possible by a low (.17) standard deviation between the subjects in the category of percent resistance. This is because cp alone is acting on the bacteria with no assistance from the individual's immune system. Environmental factors beyond the medication and bacteria adaptability like codependency and location are also key to the survival of bacteria. These factors are represented in the .17 standard deviation between individuals. Table 1 shows percent resistance highly correlates with Shannon enthropy (Q), a measurement of the structural dissimilarity of a RNA. Also the adjusted base pair distance (D) has a low p value (0.409024). Pb is the number of base pairs observed in the secondary structure and mfe is Gibbs free energy.

In Table 2 , the total resistance and total expression rows show that species with stronger secondary structures are able to be generally more expressed before, during, and after cp. The percent resistance row shows the importance of adaptation in a species' ability to flourish. Organisms are characterized by maximum presence in the gut during a regiment of cp by having a local maximum to diversity while maintaining an overall minimum secondary structure.

Table 3 shows the Pearson's Correlation coefficient and p value of 16S rRNA sequence counts for a combination of A+U juxtaposed with G+C. A high concentration of G+C, the more stable genetic structure, leads to less overall expression but a higher ability to resist the effects of cp. A+U, the less stable nucleotides, correlate with greater ability to resist cp and higher overall expression but a lesser percent of resistance. The trend is magnified only when the percentage representation of uracil and guanine doubles.

Further studying the effects of antibiotics on bacteria based on the sequential and thermodynamic characteristics of 16S rRNA, I analyzed pyroseqencing data from "Antibiotics in Feed Induce Prophages in Swine Fecal Microbiomes" (mBio). [9] In mBio, four groups of pigs were either non-medicated, given ASP250, or a low or high dose of Carbadox. Four swine were given non-medicated feed and measured 5 times over the course of 70 days. Six swine were given ASP250 for 14 days and measured twice during treatment. Four swine were given a subtherapeutic dose of Carbadox and were measured once at 14 days. Six swine were given a therapeutic dose of Carbadox for 28 days and measured twice. Three categories similar to those used for the PNAS analysis were selected. Total resistance was recorded as the amount of rarified abundances during ASP250, subtherapeutic Carbadox or therapeutic Carbadox. Non-medicated expression was recorded as the total and average non-medicated rarefied abundances. In mBio there was only one pretreatment sample and no samples taken following the cessation of treatment, so average percent resistance was recorded as the average across each swine expression during ASP250 or Carbadox divided by the average non-medicated amount. Similar to the cp analysis, there was a low standard deviation (.15) between the medicated individuals' expression. Non-medicated individuals had nearly a double standard deviation (.29) showing that without the selective pressure of antibiotics, a biome adapts to the environment of the individual.

Table 4 shows the probabilities of correlations from rarefied abundances for the three selected categories. Average percent resistance is notably statically weaker in the mBio analysis, but total resistance and non-medicated expression have statistically significant results. In Table 4, the Pb and MFE columns show a low p value for all total resistance and non-medicated categories. Pb responds similarly to the cp experiment but rRNA MFE plays a different and significant role in an organism's ability to resist and thrive in ASP250 and Carbadox. Table 5 shows that MFE has a strong negative correlation to expression during an antibiotic and expression in a non-medicated setting. A negatively correlated MFE means that unlike the cp experiment, a thermodynamically stable rRNA is selected by the ASP250 and Carbadox.

Table 6 shows that correlations and p values from A+U and G+C counts are strongly inverted from the cp experiment. Stable dinucleotide GG is selected by ASP250 and Carbadox while thermodynamically loose dinucleotide UA has a statically significant relationship to average percent resistance.

Because PNAS studies and mBio studies provide a unique view of thermodynamics of 16S rRNA during antibiotics, a third paper's data was analyzed. In JCI's "Vancomycin-resistant Enterococcus domination of intestinal microbiota is enabled by antibiotic treatment in mice and precedes bloodstream invasion in humans" mice ileum, ileum wall, cecum, and feces were analyzed in the presence of Ampicillin, Vancomycin, and an antibiotic cocktail containing Metronidazole, Neomycin, and Vancomycin (MNV). [10] Ileum and cecum was extracted from the mouse making each experiment terminal. Categories were created similar to the mBio paper. An average total expression measured the average rarefied expression of 16S reads in the untreated mice, combining ileum, ileum wall and cecum data into one measurement. Three total resistance categories contained the sum of all rarefied reads taken in the presence of Amplicin, MNV, and Vancomycin. An average percent resistance category was also created with the average expression during an antibiotic divided by the average untreated expression. Similar to the previous experiments, there was a low standard deviation between the specimen on medication (.36).

Table 7 shows the thermodynamic characteristics of JCI mice data. It is important to remember that Vancomycin is limited to gram positive bacteria and Ampicillin has a narrow ability to affect gram negative bacteria. Similar to mBio's experiment with Carbadox and ASP250, Pb and MFE play an important role in characterizing the ability of a species to resist the effects of an antibiotic. Stability again is favored for resisting all three antibiotics. Unlike mBio's results, base pairing and minimum free energy are able to characterize the percent resistance category. Table 8 shows that A+U still play an important role in characterizing an organism's ability to resist antibiotics, but G+C have a negligible effect on both. The relatively weak dimer UC measured as a percent positively correlates to total resistance and average percent resistance. The very stable dimer GG measured as a percent significantly negatively correlates to average percent resistance and total resistance.

JCI's data also contained a study on the intestinal microbiota of allogeneic hematopoietic stem cell transplantation (allo-HSCT). These patients had an underlying cancer that had been treated by chemotherapy. As the authors pointed out in JCI, it is difficult to draw conclusions from this data because of the multifold level of contributing factors to the patient's microbiota antibiotics. Each medication selects for a different set of criteria based on the method of action and most patients were on multiple antibiotics for different durations at different dosages. This was shown in the standard deviation of the samples taken during medicine being significantly larger than the PNAS and mBio (1.09). The patients with the least drug changes had a lower standard deviation among their samples. Table 9 shows that bacteria were selected from either spectrum of the thermodynamic spectrum, similar to both PNAS thermodynamically instable and mBio thermodynamically strong. The weakest dinucleotide UU positively correlates to survival of a bacterium in some patients but negatively correlates to other patients.

To further investigate the relationship of antibacterial resistance and expression, 54 features were recorded and input into a multiple linear regression (MLR). Linear regression is a method of predicting an outcome of a dependant variable given a single or multiple explanatory variables. Explanatory variables are essentially features used to describe a mathematical line that represents the dependent variables. Coefficients are modifiers of explanatory variables developed that match input data to a dependent (or training) variable. Mono and dinucleotide frequencies as well as 4 thermodynamic stability quantities and all nucleotide and thermodynamic predicators were normalized to sequence length. Three linear models were developed based on the PNAS, mBio, and JCI data. When the linear model was asked to predict the simple percent of resistance of PNAS, it was accurate to 20% of the actual resistance in 78% of the rRNA. It was accurate to 10% of actual resistance for 57% of all responses. The binomial distribution of 10% (3/10) and 20% (1/2) accuracy for 2827 samples is 0. Table 10 shows that the ability to predict resistance was generalized across the families of the bacteria kingdom. The large Firmicutes and Bacteroidetes families both performed well, as evidenced by the average accuracy of 14%. This shows that the human gut's microbiota's ability to resist antibiotics can be quantified by rRNA sequential and thermodynamic characteristics. To increase the specificity of the linear model, the rRNA of the PNAS samples were shuffled using a simple mononuclide shuffle and added to the training set. The random sequences were identified by negative number or numbers with a percent less than .01 in 81% of the time. 99% of all shuffled sequences properly identified as having a percent resistance of less than .2. This marginally decreased the accuracy of the MLR to 75%.

A separate multiple linear regression model was built based on the mBio data with the same features as the PNAS paper. The predicted variable had to be altered to fit the data because there was no preceding or continuing data after treatment from the experiment. Instead a variable was selected similar to percent resistance (the average amount of resistance divided by the sum of the average non-medicated and average medicated samples). This number is useful because like percent resistance it tells what resistance to expect given an initial amount. The sub-therapeutic Carbox MLR performed similar to the cp experiment yielded an accuracy of 20% for 78% of the data. The therapeutic Carbox and ASP250 had a lower accuracy of 20% for 56% and 54%, respectively, of the samples. Table 11 shows that the ability to predict the ASP250, sub therapeutic and therapeutic Carbox. Firmacutes performed with an average accuracy of 25%, 16% and 25% for ASP250, Subtherapeutic and Therapeutic Carbox, respectively. Bacteroidetes likewise performed with an average accuracy of 23%, 15% and 21%. When mononuclide shuffled reads were added to increase specificity, the average accuracy increased in ASP250 to 62%. Subtherapeutic Carbox likewise increased to 76% and therapeutic Carbox increased to 61%. Randomly shuffled sequences were correctly identified by the models as having a percent resistance of less than .01 82% of the time. A third MLR was created for the JCI data but it did not perform as accurately as the PNAS data or mBio data, but the results were still statistcally significant based on the Binomial distribution (see Table 12). In the Ampicillin, Vancomycin and MNV experiment the MLR was accurate to 20% of the actual resistance in 38%, 45% and 17% of the rRNA. One explanation for the lower accuracy is the lower depth. There were significantly fewer Firmacutes (561) and Bacteroidetes (5) in the JCI data when compared with the PNAS (2235, 478). Also the medications used in the JCI experiment had a narrower spectrum of influence. Ampicillin only effects some gram negatives while Vancomycin affects only gram positives, but the MNV cocktail should effect both gram positive and negative bacteria.

16S rRNA sequential characteristics extend beyond taxonomy and are able to guide us in a species-based analysis of antibiotic resistance. 16S rRNA would not provide information on strain-based resistance because two strains share the same 16S rRNA.

The MLRs were used to make predictions concerning the percent resistance from 32 families found in the gut from the SILVA database. A contentious of each MLR was made and 172 species were found to have a high probability of antibiotic resistance. The results included Eubacterium brachy, a chest infection which requires multiple days of penicillin [26] and Acanthamoeba, which can cause blindness when present in the eye and is immune to many contact disinfectants [27]. Also included was Prevotella timonensis, which showed immunity to penicillin in a study [28]. Additionally Streptococcus, the cause of strep throat, was isolated. An uncultured Fusobacterium was also identified as potentially antibiotic resistant. Some Fusobacterium infections are called Lemierre's syndrome, a complication of strep throat in which the lesions caused by strep throat become infected, ultimately causing severe pneumonia [29]

The Antibiotic Resistance Genes Database (ARGB) contains a list of genus and species known to have antibiotic resistance. Genera, which were predicted to have at least 10% percent resistant in cp, ASP250 and Carbadox experiments were cross correlated with the ARGB. 21 Genera were validated by ARDB and are shown in Table 13. Most Genera were resistant to tetracycline. Mutations in the 16S region are known to cause resistance to tetracycline. 9 of the 21 were also resistant to bacitracin, an ingredient in Neosporin [30] [31].

**Acknowledgments**

I thank Les Dethlefsen for providing his 16S rRNA data and for help with conceptualizing the comparison of samples. I also would like to thank Heather K. Allen for providing her data from mBio and Carles Ubeda for providing data from his JCI experiments. Finally, I thank the NCBI SRA for hosting said data.

Table 1. PNAS probabilities of correlations to thermodynamic metrics.

| | Pb | Npb | Mfe | Nmfe | Q | NQ | D | ND |
|---|---|---|---|---|---|---|---|---|
| Total resistance | 0.000408 | 3.46E-06 | 0.183175 | 0.000467 | 0.15929 | 0.242598 | 0.13884 | 0.222275 |
| Average percent resistance | 0.000791 | 4.72E-06 | 0.000116 | 1.49E-05 | 1.98E-05 | 4.71E-06 | 8.99E-05 | 2E-05 |
| Total expression | 0.000196 | 1.03E-05 | 0.418445 | 0.026868 | 0.458746 | 0.380076 | 0.409024 | 0.415018 |

Table 2. PNAS correlations and p value of correlations to normalized number of base pairs observed in the secondary structure

| | Npb | |
|---|---|---|
| | Pearson's $r$ | P value |
| Total resistance | 0.075629 | 1.4E-05 |
| Average percent resistance | -0.0923 | 1.55E-07 |
| Expression | 0.090124 | 2.93E-07 |

Table 3. PNAS Pearson's correlation and p value for total amount of A+U, G+C, UU and GG

| | A+U | | G+C | | UU | | GG | |
|---|---|---|---|---|---|---|---|---|
| | Pearson's $r$ | p value | Pearson's $r$ | p value | Pearson's $r$ | p value | Pearson's $r$ | p value |
| Total resistance | 0.105349 | 3.55E-09 | -0.04589 | 0.005941 | 0.141452 | 3.35E-15 | -0.06075 | -0.06075 |
| Average percent resistance | -0.03071 | 0.046166 | -0.01214 | 0.252917 | -0.06019 | 0.000481 | 0.02179 | 0.02179 |
| Total expression | 0.088504 | 5.9E-07 | -0.02253 | 0.108477 | 0.113056 | 2.57E-10 | -0.04411 | -0.04411 |

Table 4. mBio probabilities of correlation of thermodynamics to abundance categories.

|  |  | Pb | Npb | Mfe | Nmfe | Q | NQ | D | ND |
|---|---|---|---|---|---|---|---|---|---|
| Total resistance | asp250 | 7.41E-08 | 0.000596 | 1.05E-10 | 8.15E-07 | 0.06106 | 0.230059 | 0.026053 | 0.147744 |
|  | subther | 8.56E-07 | 0.000261 | 1.13E-08 | 6.28E-06 | 0.357704 | 0.368773 | 0.222416 | 0.496391 |
|  | Ther | 0.00177 | 0.007165 | 2.08E-07 | 4.17E-07 | 0.487241 | 0.349558 | 0.34493 | 0.498942 |
| Average percent resistance | asp250 | 0.2401 | 0.477282 | 0.359684 | 0.136471 | 0.471275 | 0.424148 | 0.490238 | 0.45235 |
|  | subther | 0.215492 | 0.23536 | 0.008 | 0.002916 | 0.47039 | 0.418376 | 0.385847 | 0.434622 |
|  | Ther | 0.085062 | 0.346001 | 0.13053 | 0.476685 | 0.216238 | 0.337505 | 0.135454 | 0.246614 |
| Non-medicated expression | nonmed | 6.17E-08 | 4.04E-06 | 2.46E-09 | 1.87E-06 | 0.343282 | 0.391542 | 0.218599 | 0.493971 |
|  | nonmedaverage | 7.19E-09 | 1.69E-06 | 4.2E-10 | 1.15E-06 | 0.212889 | 0.473431 | 0.108238 | 0.337809 |

Table 5. mBio MFE characterizes antibiotic adaptation and overall expression.

|  |  | MFE | |
|---|---|---|---|
|  |  | Pearson's $r$ | p value |
| Total resistance | asp250 | -0.12364 | 1.05E-10 |
|  | Subther | -0.10891 | 1.13E-08 |
|  | Ther | -0.09864 | 2.08E-07 |
| Average percent resistance | asp250 | 0.007019 | 0.359684 |
|  | Subther | 0.047032 | 0.008 |
|  | Ther | -0.02195 | 0.13053 |
| Total non-medicated expression | Nonmed average | -0.11946 | 4.2E-10 |

Table 6. mBio Pearson's correlations and p values for total amount of A+U, G+C, UU and GG

|  |  | A+U | | G+C | | UA | | GG | |
|---|---|---|---|---|---|---|---|---|---|
|  |  | Pearson's *r* | p value | Pearson's *r* | p value | Pearson's *r* | p value | Pearson's *r* | p value |
| Total resistance | asp250 | -0.00254 | 0.44834 | 0.116055 | 1.25E-09 | -0.00812 | 0.338852 | 0.100384 | 1.29E-07 |
|  | subther | 0.003916 | 0.420561 | 0.093139 | 8.84E-07 | 0.015192 | 0.218369 | 0.092743 | 9.78E-07 |
|  | ther | -0.03304 | 0.045322 | 0.0842 | 7.88E-06 | -0.03151 | 0.053306 | 0.104315 | 4.29E-08 |
| Average percent resistance | asp250 | 0.024058 | 0.109029 | -0.00173 | 0.464656 | 0.032688 | 0.047086 | -0.01676 | 0.195408 |
|  | subther | 0.044442 | 0.011419 | -0.05516 | 0.002357 | 0.080918 | 1.67E-05 | -0.0605 | 0.000968 |
|  | ther | 0.035101 | 0.036138 | 0.014517 | 0.228687 | 0.039448 | 0.021685 | 0.002101 | 0.457167 |
| Total non-medicated expression | Nonmed avg | -0.00149 | 0.469674 | 0.109066 | 1.07E-08 | -0.01895 | 0.166022 | 0.098584 | 2.11E-07 |

Table 7. JCI Pearson's correlations and p values of number of base pairs observed in secondary structure of Pb and MFE

|  |  | Pb | | MFE | |
|---|---|---|---|---|---|
|  |  | Pearson's *r* | p value | Pearson's *r* | p value |
| Total resistance | Ampicillin | 0.132085 | 0.000245 | -0.08994 | 0.008935 |
|  | Vancomycin | 0.091106 | 0.008219 | -0.0826 | 0.014842 |
|  | MVN | 0.120199 | 0.000762 | -0.08983 | 0.009009 |
| Average percent resistance | Ampicillin | 0.147611 | 4.81E-05 | -0.12434 | 0.000519 |
|  | Vancomycin | 0.174791 | 1.84E-06 | -0.18255 | 6.57E-07 |
|  | MVN | 0.173781 | 2.1E-06 | -0.19521 | 1.11E-07 |
| Total non-medicated expression | untreated average | 0.061816 | 0.051983 | -0.02676 | 0.240956 |

Table 8. JCI Pearson's correlations and p values for total amount of A+U, G+C, UU and GG

|  |  | A+U | | G+C | | UC | | GG | |
|---|---|---|---|---|---|---|---|---|---|
|  |  | Pearson's r | p value | Pearson's r | p value | Pearson's r | p value | Pearson's r | p value |
| Total resistance | Ampicillin | 0.093382 | 0.006962 | 0.034653 | 0.181184 | 0.071864 | 0.029323 | -0.06792 | 0.451644 |
|  | Vancomycin | 0.054055 | 0.077591 | 0.038467 | 0.155964 | 0.100866 | 0.003939 | -0.00462 | 0.147026 |
|  | MVN | 0.082307 | 0.015137 | 0.043432 | 0.126767 | 0.093491 | 0.006906 | -0.03991 | 0.036989 |
| Average percent resistance | Ampicillin | 0.12024 | 0.000759 | 0.035066 | 0.178338 | 0.090578 | 0.008539 | -0.11979 | 0.000791 |
|  | Vancomycin | 0.113858 | 0.001343 | 0.053345 | 0.080342 | 0.141732 | 9.08E-05 | -0.08877 | 0.009716 |
|  | MVN | 0.115643 | 0.001148 | 0.043638 | 0.125641 | 0.146904 | 5.19E-05 | -0.10026 | 0.004129 |
| Total non-medicated expression | untreated average | 0.05476 | 0.074929 | 0.030879 | 0.208509 | 0.003146 | 0.46706 | -0.03355 | 0.188892 |

Table 9. JCI allo-HSCT Pearson's correlations and p values for total amount of UU

| | UU | |
|---|---|---|
| Patient Average Expression | Pearson's r | p value |
| A | -0.15232 | 0.014817 |
| B | 0.140358 | 0.022625 |
| C | 0.117261 | 0.04743 |
| D | -0.15141 | 0.015316 |
| E | 0.126492 | 0.035709 |

Table 10. PNAS accuracy of predictions by families

| | Taxonomy Name | Total number of rRNA | 10% accuracy | 20% accuracy | Average accuracy | Average error | Binomial 10% | Binomial 20% |
|---|---|---|---|---|---|---|---|---|
| Kingdom | Bacteria | 2827 | 1425 | 2198 | 0.138336 | 0.715457 | 0 | 0 |
| Family | RF3 | 1 | 1 | 1 | 0.026151 | 0.026306 | 0.3 | 0.5 |
| | Tenericutes | 23 | 12 | 19 | 0.10424 | 0.24925 | 0.014208075 | 0.001056 |
| | Cyanobacteria | 6 | 4 | 6 | 0.095893 | 0.552499 | 0.059535 | 0.015625 |
| | Bacteroidetes | 478 | 228 | 360 | 0.142242 | 2.888332 | 1.96337E-16 | 5.62E-30 |
| | Synergistetes | 1 | 1 | 1 | 0.047813 | 0.011356 | 0.3 | 0.5 |
| | Firmicutes | 2235 | 1132 | 1745 | 0.138862 | 0.251931 | 0 | 0 |
| | Fusobacteria | 1 | | 1 | 0.1497 | 0.04366 | 0.7 | 0.5 |
| | Proteobacteria | 29 | 23 | 26 | 0.086284 | 0.175718 | 5.26125E-08 | 6.81E-06 |
| | CandidatedivisionTM7 | 2 | | | 0.203213 | 0.068636 | 0.49 | 0.25 |
| | Actinobacteria | 37 | 19 | 26 | 0.137904 | 0.044192 | 0.003344797 | 0.006221 |
| | Lentisphaerae | 3 | | 3 | 0.134658 | 0.066854 | 0.343 | 0.125 |
| | Verrucomicrobia | 10 | 4 | 9 | 0.109075 | 6.270987 | 0.200120949 | 0.009766 |
| | Acidobacteria | 1 | 1 | 1 | 0.025956 | 0.000844 | 0.3 | 0.5 |

Table 11. mBio accuracy of predictions by families

| | | | ASP 250 | | | SubTher | | | Ther | | |
|---|---|---|---|---|---|---|---|---|---|---|---|
| | Taxonomy Name | Total number of rRNA | Average accuracy | Binomial 10% | Binomial 20% | Average accuracy | Binomial 10% | Binomial 20% | Average accuracy | Binomial 10% | Binomial 20% |
| Kingdom | Bacteria | 2481 | 0.248564 | 0.016408 | 2.02E-06 | 0.168523 | 0 | 0 | 0.232437 | 0.007889 | 4.08E-11 |
| Family | RF3 | 2 | 0.458346 | 0.49 | 0.5 | 0.184207 | 0.49 | 0.5 | 0.554129 | 0.49 | 0.25 |
| | Tenericutes | 73 | 0.237729 | 0.099993 | 0.092439 | 0.280823 | 0.003475 | 0.020137 | 0.358191 | 7.53E-05 | 0.001254 |
| | Cyanobacteria | 41 | 0.317204 | 0.11076 | 0.11126 | 0.375288 | 0.12615 | 0.046874 | 0.395855 | 0.048465 | 0.008013 |
| | Planctomycetes | 3 | 0.140878 | 0.441 | 0.125 | 0.053357 | 0.027 | 0.125 | 0.143149 | 0.441 | 0.375 |
| | Synergistetes | 1 | 0.082512 | 0.3 | 0.5 | 0.061252 | 0.3 | 0.5 | 0.019074 | 0.3 | 0.5 |
| | Firmicutes | 1605 | 0.255339 | 0.013898 | 0.000111 | 0.162195 | 0 | 0 | 0.22746 | 0.005963 | 8.57E-10 |
| | Elusimicrobia | 4 | 0.308899 | 0.2401 | 0.25 | 0.435062 | 0.2401 | 0.375 | 0.169051 | 0.2646 | 0.25 |
| | Fusobacteria | 2 | 0.23118 | 0.49 | 0.25 | 0.1579 | 0.42 | 0.5 | 0.2096 | 0.49 | 0.5 |
| | Bacteroidetes | 654 | 0.230957 | 0.013604 | 0.001361 | 0.154454 | 2.83E-45 | 8.21E-54 | 0.213987 | 0.006066 | 7.12E-07 |
| | Proteobacteria | 19 | 0.192304 | 0.191639 | 0.05175 | 0.24693 | 0.152529 | 0.022179 | 0.395266 | 0.086947 | 0.007393 |
| | CandidatedivisionTM7 | 2 | 0.141668 | 0.49 | 0.25 | 0.041356 | 0.09 | 0.25 | 0.036763 | 0.09 | 0.25 |
| | Fibrobacteres | 5 | 0.192572 | 0.16807 | 0.3125 | 0.139283 | 0.3087 | 0.15625 | 0.363807 | 0.36015 | 0.15625 |
| | Actinobacteria | 4 | 0.206622 | 0.4116 | 0.375 | 0.198609 | 0.4116 | 0.25 | 0.216999 | 0.2401 | 0.375 |
| | Lentisphaerae | 20 | 0.231103 | 0.191639 | 0.120134 | 0.116674 | 0.001018 | 0.001087 | 0.220939 | 0.164262 | 0.120134 |
| | Verrucomicrobia | 4 | 0.242314 | 0.2401 | 0.25 | 0.323671 | 0.2401 | 0.25 | 0.484761 | 0.2401 | 0.0625 |
| | Spirochaetes | 42 | 0.262536 | 0.119639 | 0.116823 | 0.198674 | 0.067851 | 0.005802 | 0.237984 | 0.119639 | 0.022436 |

Table 12. JCI accuracy of predictions by families

|  | | | AMP | | | Vancomycin | | | MNV | | |
|---|---|---|---|---|---|---|---|---|---|---|---|
|  | Taxonomy Name | Total number of rRNA | Average accuracy | Binomial 10% | Binomial 20% | Average accuracy | Binomial 10% | Binomial 20% | Average accuracy | Binomial 10% | Binomial 20% |
| Kingdom | Bacteria | 661 | 0.30361 | 3.57E-06 | 2.96E-08 | 0.284937 | 8.07E-05 | 0.002658 | 0.50405 | 1.52E-18 | 1.42E-30 |
| Family | Verrucomicrobia | 2 | 0.601721 | 0.42 | 0.5 | 0.423359 | 0.49 | 0.5 | 0.923359 | 0.49 | 0.25 |
|  | Cyanobacteria | 5 | 0.380182 | 0.16807 | 0.03125 | 0.313716 | 0.16807 | 0.3125 | 0.307869 | 0.16807 | 0.3125 |
|  | Proteobacteria | 27 | 0.424958 | 0.004236 | 0.006616 | 0.357437 | 0.038906 | 0.03492 | 0.645218 | 6.57E-05 | 2.62E-06 |
|  | Chloroflexi | 1 | 0.623925 | 0.7 | 0.5 | 0.13029 | 0.7 | 0.5 | 0.86971 | 0.7 | 0.5 |
|  | Firmicutes | 561 | 0.288112 | 0.00029 | 5.31E-05 | 0.267278 | 0.007556 | 0.019306 | 0.48864 | 3.54E-12 | 1.37E-21 |
|  | Acidobacteria | 1 | 0.213764 | 0.7 | 0.5 | 0.261073 | 0.7 | 0.5 | 0.261073 | 0.7 | 0.5 |
|  | Actinobacteria | 56 | 0.381257 | 0.003421 | 2.62E-05 | 0.434102 | 2.62E-05 | 8.06E-05 | 0.633635 | 4.62E-06 | 5.3E-11 |
|  | Bacteroidetes | 5 | 0.173041 | 0.1323 | 0.3125 | 0.135597 | 0.36015 | 0.03125 | 0.134753 | 0.36015 | 0.03125 |
|  | Nitrospirae | 1 | 0.903102 | 0.7 | 0.5 | 0.469787 | 0.7 | 0.5 | 0.469787 | 0.7 | 0.5 |
|  | Tenericutes | 2 | 0.260503 | 0.49 | 0.5 | 0.242651 | 0.49 | 0.5 | 0.242651 | 0.49 | 0.5 |

Table 13. Select genera from predictions mapped to ARDB with select drug resistances

| Genus | tetracycline | bacitracin | chloramphenicol | lincosamide | macrolide | streptomycin | vancomycin |
|---|---|---|---|---|---|---|---|
| Finegoldia | x | | | | | | |
| Eubacterium | x | x | x | | | | x |
| Lactococcus | x | | x | x | | x | |
| Streptococcus | x | | x | x | x | | x |
| Faecalibacterium | x | | | | | | |
| Ruminococcaceae | x | | | | | | |
| Prevotell | x | | | | | | |
| Acidaminococcus | x | x | | | | | |
| Anaerotruncus | x | | | | | | |
| Selenomonas | x | x | | | | | |
| Anaerococcus | x | x | | | | | |
| Enterococcus | x | x | x | x | x | x | x |
| Anaerostipes | x | x | | | | | |
| Lactobacillus | x | x | x | x | x | | |
| Pediococcus | | | | x | | x | |
| Coprococcus | x | x | | | | | |
| Clostridium | x | x | x | x | x | x | x |
| Ruminococcus | | | | | | | x |
| Dorea | x | | | | | x | |
| Butyrivibrio | x | | | | | | |
| Granulicatella | | | | | x | | |
| Total count | 18 | 9 | 6 | 6 | 5 | 5 | 5 |